\documentclass[twocolumn,showpacs,superscriptaddress,amsmath,amssymb,nofootinbib,prl]{revtex4-2}
\usepackage{graphicx}
\usepackage{subfigure}
\usepackage{units}
\usepackage{multirow}
\usepackage{epsfig}
\usepackage{dcolumn}
\usepackage{booktabs}
\usepackage{appendix}
\usepackage{bm}
\usepackage{overpic}
\usepackage{color}
\usepackage{threeparttable}
\usepackage{lineno}
\usepackage{float}
\usepackage{epstopdf}
\usepackage[colorlinks,linkcolor=blue,anchorcolor=blue,citecolor=blue,urlcolor=blue]{hyperref}
\usepackage{ulem}

\def\degree#1{{#1}^{\circ}}

\begin{document}
\title{\bf\boldmath Search for the decay \texorpdfstring{$D^{0} \to \pi^{0} \nu \bar{\nu}$}{D0->pi0nunubar}}
\date{\it \small \bf \today}
\author{\small
M.~Ablikim$^{1}$, M.~N.~Achasov$^{10,b}$, P.~Adlarson$^{66}$, S. ~Ahmed$^{14}$, M.~Albrecht$^{4}$, R.~Aliberti$^{27}$, A.~Amoroso$^{65A,65C}$, M.~R.~An$^{31}$, Q.~An$^{62,48}$, X.~H.~Bai$^{56}$
% ...
\\
\vspace{0.2cm}
(BESIII Collaboration)\\
\vspace{0.2cm} {\it
$^{1}$ Institute of High Energy Physics, Beijing 100049, People's Republic of China\\
% ...
$^{4}$ Bochum Ruhr-University, D-44780 Bochum, Germany\\
% ...
$^{10}$ G.I. Budker Institute of Nuclear Physics SB RAS (BINP), Novosibirsk 630090, Russia\\
% ...
$^{14}$ Helmholtz Institute Mainz, Staudinger Weg 18, D-55099 Mainz, Germany\\
% ...
$^{27}$ Johannes Gutenberg University of Mainz, Johann-Joachim-Becher-Weg 45, D-55099 Mainz, Germany\\
% ...
$^{31}$ Liaoning Normal University, Dalian 116029, People's Republic of China\\
% ...
$^{48}$ State Key Laboratory of Particle Detection and Electronics, Beijing 100049, Hefei 230026, People's Republic of China\\
% ...
$^{56}$ University of Jinan, Jinan 250022, People's Republic of China\\
% ...
$^{62}$ University of Science and Technology of China, Hefei 230026, People's Republic of China\\
% ...
$^{65}$ University of Turin and INFN, (A)University of Turin, I-10125, Turin, Italy; (B)University of Eastern Piedmont, I-15121, Alessandria, Italy; (C)INFN, I-10125, Turin, Italy\\
$^{66}$ Uppsala University, Box 516, SE-75120 Uppsala, Sweden\\
% ...
$^{b}$ Also at the Novosibirsk State University, Novosibirsk, 630090, Russia\\
% ...
}\vspace{0.4cm}}

\begin{abstract}
We present the first experimental search for the rare charm decay $D^{0} \to \pi^{0} \nu \bar{\nu}$. It is based on an $e^+e^-$ collision sample consisting of $10.6\times10^{6}$ pairs of $D^0\bar{D}^0$ mesons collected by the BESIII detector at $\sqrt{s}$=3.773~GeV, corresponding to an integrated luminosity of 2.93~fb$^{-1}$. A data-driven method is used to ensure the reliability of the background modeling. No significant  $D^{0} \to \pi^{0} \nu \bar{\nu}$ signal is observed in data and an upper limit of the branching fraction is set to be $2.1\times 10^{-4}$ at the 90$\%$ confidence level. This is the first experimental constraint  on charmed-hadron decays into dineutrino final states. 
\end{abstract}

\maketitle

Flavour changing neutral current (FCNC) transitions in the Standard Model (SM) are highly suppressed by the Glashow-Iliopoulos-Maiani$\,$(GIM) mechanism~\cite{Glashow:1970gm}. GIM suppression is more effective for the charm sector compared to the down-type quarks in the bottom and strange sectors. This suppression is responsible for the relatively small size of charm mixing and $C\!P$ violation in the charm system~\cite{Asner:2008nq,Saur:2020rgd,Wilkinson:2021tby}. Therefore, FCNC processes involving $D$ decays into charged lepton pairs are often totally overshadowed by long distance contributions~\cite{Cappiello:2012vg,Aaij:2017iyr}. However, for $D$ FCNC decays into final states involving dineutrinos, such as $D^0\to \pi^0 \nu \bar{\nu}$, long-distance contributions become insignificant and the short-distance contributions from $Z$-penguin and box diagrams dominate, resulting in the branching fraction at the level of $10^{-15}$ in SM~\cite{Burdman:2001tf}. That makes $D$ FCNC decay involving dineutrinos a unique and clean probe to study the $C\!P$ violation in the charm sector~\cite{2012JHEP03021B} and search for new physics beyond  SM~\cite{Paul:2011ar}. Such new physics effects can enhance the branching fraction largely above the tiny values in SM~\cite{Bause:2020xzj,Fajfer:2021woc}. 

Recently, LHCb reported evidence for the breaking of lepton universality in bottom-quark FCNC decays to charged dielectrons and dimuons with a significance of 3.1$\sigma$~\cite{LHCb:2021trn}, which suggests the possible presence of new physics contributions in the lepton sector~\cite{European}. For instance, a popular proposal of leptoquarks beyond SM, which have different interaction strengths with the different types of leptons, can naturally accommodate the lepton universality breaking~\cite{Dorsner:2016wpm,Bause:2020xzj}.
A complementary study on the FCNC decay $D^0\to \pi^0 \nu \bar{\nu}$ is necessary to test the flavour structure of leptons given its extremely suppressed decay rate~\cite{Burdman:2001tf} in SM. In the scenario of leptoquarks, the branching fraction can be enhanced up to $9.7\times10^{-4}$ with inclusion of the sterile neutrinos~\cite{Faisel:2020php}.

Charm decays with a neutrino pair in the final state have never been searched for experimentally. Taking advantage of the clean environment of threshold production, the first search for $D$ meson FCNC decays involving dineutrinos is performed based on the 2.93 $\mbox{fb$^{-1}$}$ $e^+e^-$ annihilation data sample collected at $\sqrt{s}=3.773\,\mathrm{GeV}$ by the BESIII detector~\cite{3770luminosity}. 
This sample consists of $10.6\times10^{6}$ pairs of $D^0\bar{D}^0$ mesons and $8.3\times10^{6}$ pairs of $D^+D^-$ mesons ~\cite{crosssection}. 
Decays of $D^0$ mesons into dineutrino final states are more suitable for study than those of their charged counterpart, as $D^+$ decays of this nature suffer from irreducible contamination from the process $D^+\to\tau^+\nu_{\tau}$, where  the $\tau$ lepton decays into mesons and a neutrino. Consequently, this letter reports a search for  the  FCNC decay $D^{0} \to \pi^{0} \nu \bar{\nu}$. For this decay process, a reliable modeling of the background contributions is crucial. This requirement is best achieved through  a sophisticated data-driven method. 
About one third of the full data sample is used to validate the data-driven analysis procedure, and the final results are obtained by unblinding the total data set after the analysis strategy is fixed.

Details about the BESIII detector design and performance are given  in Ref.~\cite{Ablikim:2009aa}. A generic Monte Carlo (MC) simulated data sample, described in Ref.~\cite{Ablikim:2018gro}, is employed to identify sources of background contributions. 
The signal process $D^0\to \pi^0 \nu \bar{\nu}$ is simulated following the model described in Ref.~\cite{Bause:2020xzj}, incorporating associated $D\to \pi$ form factors in the form of a single-pole model with parameters  measured in Ref.~\cite{Ablikim:2015ixa}.


Since the center-of-mass energy of 3.773 GeV is just above the $D\bar{D}$ mass threshold, the $D^{0}$ and $\bar{D}{}^{0}$ mesons are produced almost at rest without any additional hadrons in the event. Hence, it is straightforward to use a double-tag approach~\cite{Baltrusaitis:1985iw,Li:2021iwf} to measure the absolute branching fraction of $D^0\to \pi^0 \nu \bar{\nu}$, $\mathcal{B}_{\rm sig}$, according to 
\begin{equation}\label{equ:bsigsDT}
\mathcal{B}_{\rm{sig}}=\frac{N_{\rm{sig}}}{\mathcal{B}_{\pi^0 \to \gamma\gamma}\sum_{\alpha}N_{\rm{tag}}^{\alpha}\epsilon_{\rm{tag,sig}}^{\alpha}/\epsilon_{\rm{tag}}^{\alpha}}.
\end{equation}
Here, $\alpha$ represents the different single-tag modes, $N_{\rm{tag}}^{\alpha}$ is the single-tag yield for tag mode $\alpha$, $N_{\rm{sig}}$ is the sum of the double-tag yields from all single-tag modes, $\epsilon_{\rm{tag}}^{\alpha}$ and $\epsilon_{\rm{tag,sig}}^{\alpha}$ refer to the corresponding single-tag efficiency and the double-tag efficiency, respectively, for the tag-mode $\alpha$ as determined by MC simulations, and $\mathcal{B}_{\pi^0 \to \gamma\gamma}$ is the branching fraction of $\pi^0 \to \gamma\gamma$. In this approach, the systematic uncertainties arising from the single-tag reconstruction are canceled out. Throughout this Letter, charge-conjugate modes are always implied.

A detailed description of the selection criteria for charged and neutral particle candidates is provided in Ref.~\cite{Ablikim:2018gro}. The single-tag $\bar{D}{}^{0}$ candidates are reconstructed by appropriate combinations of the charged tracks and $\pi^{0}$ candidates forming the three hadronic channels: $K^{+} \pi^{-}$, $K^{+} \pi^{-} \pi^{0}$ and $K^{+} \pi^{-} \pi^{+} \pi^{-}$. The candidates are selected using two variables calculated in the $e^+e^-$ center-of-mass frame: the beam-constrained mass $M_{\rm{BC}}=\sqrt{E_{\rm{beam}}^2/c^{4} - |\textbf{p}_{\bar{D}^0}|^2/c^{2}}$ and the energy difference $\Delta E = E_{\bar{D}{}^{0}} - E_{\rm{beam}}$. Here, $E_{\rm{beam}}$ is the energy of the electron beam and $E_{\bar{D}{}^{0}}$ and $\textbf{p}_{\bar{D}^0}$ are the total energy and momentum, respectively, of all final-state particles of the $\bar{D}{}^{0}$ candidate. If there are multiple combinations for each tag mode in one event, the candidate with the smallest $\left | \Delta E \right |$ is chosen. The single-tag yield $N_{\rm{tag}}^{\alpha}$  of each tag mode is obtained from a fit to the $M_{\rm{BC}}$ distribution in data and the  efficiency $\epsilon_{\rm{tag}}^{\alpha}$ is determined from the generic MC sample following the method in Ref.~\cite{Ablikim:2018gro}.

\begin{figure}[h!]
\centering
\subfigure[]
{\includegraphics[width=\linewidth]{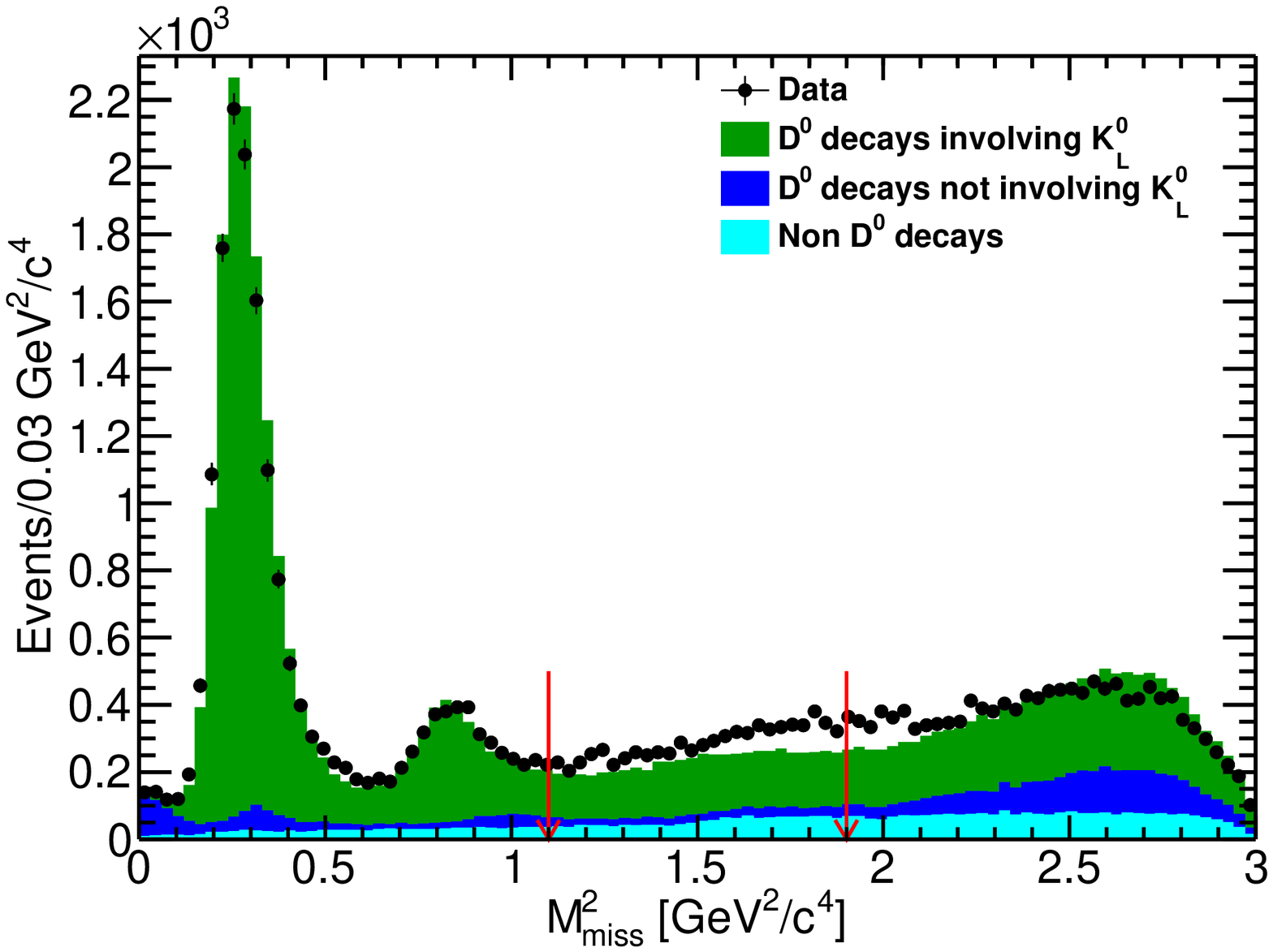}\label{fig:mm2}}
\subfigure[]
{\includegraphics[width=\linewidth]{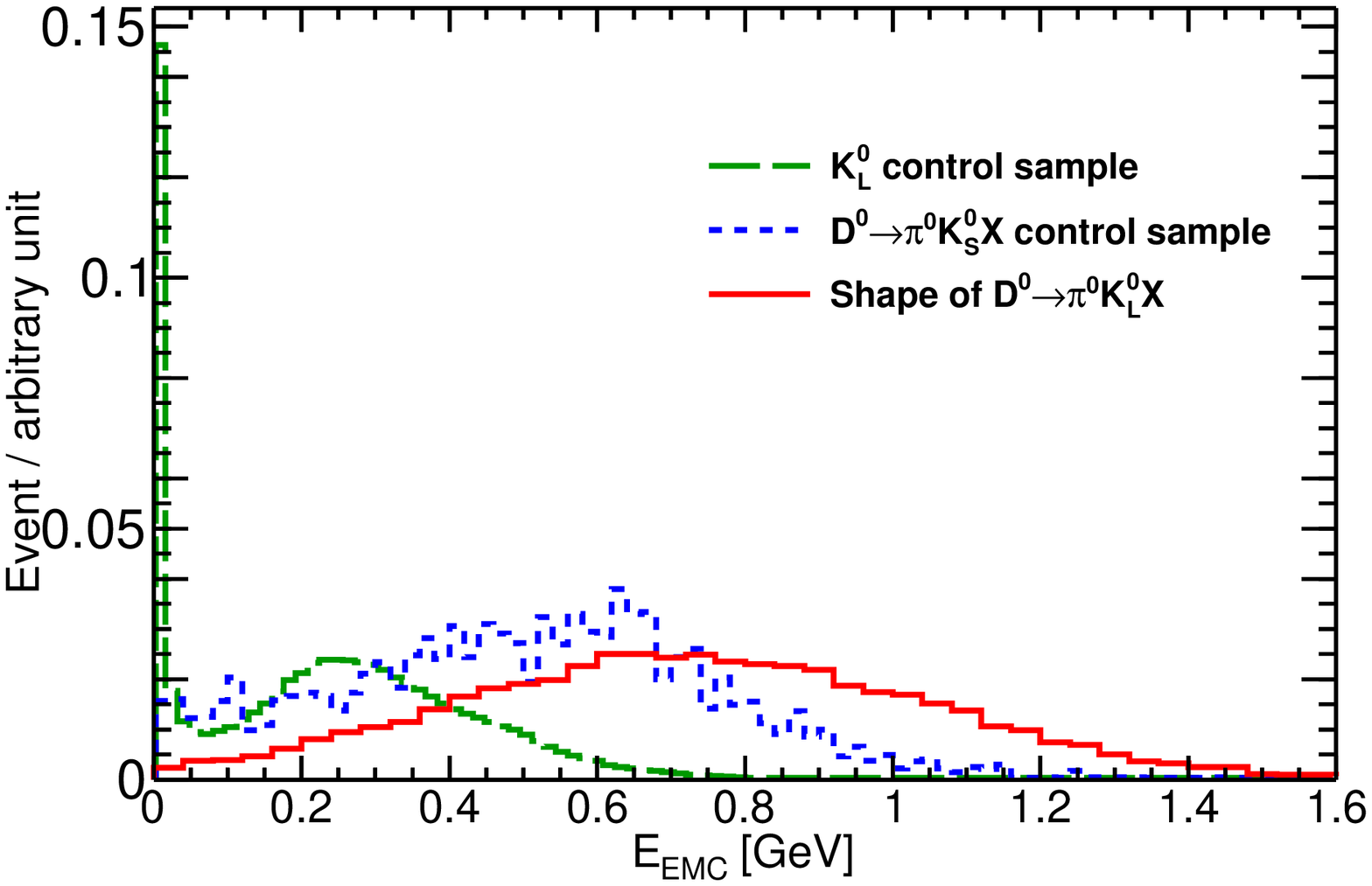}\label{fig:klpi0xshape}}
\caption{(a) $M_{\rm{miss}}^2$ distribution in data and generic MC sample. The black dots with uncertainties represent data and the filled areas show the
background components estimated from generic MC sample. The green, blue
and cerulean filled areas represent the background components of $D^0$
decays with a $K_L^0$ meson in the final state, $D^0$ decays without
$K_L^0$ meson in the final state and non $D^0$ decays, respectively.  The vertical arrows indicate the selected region.
 (b) Distributions of $E_{\rm EMC}$, the summed calorimeter energy unassociated with signal and tag decays. The red solid line shows the distribution of the $D^0\to \pi^0 K_{L}^0 X$ background. Energy deposit of $K_{L}^0$ and $X$ obtained from the corresponding control samples are marked as green long-dashed and blue dashed lines, respectively. All the distributions are normalized to unity.}
\end{figure}

After a single-tag $\bar{D}{}^{0}$ meson is identified, we reconstruct the signal $D^{0}$ decay recoiling against the single-tag $\bar{D}{}^{0}$ meson. As the signal final-state particles are all neutral, we require there  are no charged tracks in addition to those of the tagged $\bar{D}{}^{0}$ meson. Furthermore, it is required that only one $\pi^{0}$ candidate can be formed from any pairs of photon candidates not used to reconstruct the tagged $\bar{D}{}^{0}$ meson. The invariant mass of two photons for the $\pi^{0}$ candidate must satisfy  $M_{\gamma\gamma}$ $\in$ (0.115, 0.150) $\,\mathrm{GeV}/c^2$ and pass the kinematic fit constraining $M_{\gamma\gamma}$ to the nominal $\pi^{0}$ mass. The $\chi^2$ of the kinematic fit is required to be less than 20 to improve background suppression. By studying the generic MC sample~\cite{Zhou:2020ksj}, we find the surviving backgrounds mainly come from $D^0$ decays involving a $K_{L}^{0}$ meson. Figure~\ref{fig:mm2} presents the distribution of the recoil mass $M_{\rm{miss}}^2 c^2=(p_{e^+e^-} - p_{\bar D^0} - p_{\pi^0})^2$, where the four-momenta $p_{e^+e^-}$, $p_{\bar D^0}$, and $p_{\pi^0}$ of the initial $e^+e^-$ system, the tag $\bar D^0$, and the signal-side $\pi^0$, respectively, are evaluated in the $e^+e^-$ center-of-mass system.  Two peaks around 0.25 and 0.80 $\mathrm{GeV^2}/c^4$ are prominent, which correspond to the processes $D^0\to K_{L}^{0} \pi^0$ and $\bar{K}^{\ast}(892)^0\pi^0$, respectively. To remove these background contributions, only events 
with $M_{\rm{miss}}^2 \in (1.1, 1.9)\,\mathrm{GeV^2}/c^4$ are retained. The upper bound is applied to suppress the backgrounds with multiple charged and neutral soft pions, which give rise to a broad bump near the kinematic limit of 3.0 $\mathrm{GeV^2}/c^4$. After imposing all the above requirements on the signal MC samples,
the ratios of $\epsilon_{\rm{tag,sig}}^{\alpha}/\epsilon_{\rm{tag}}^{\alpha}$ in Eq.~\eqref{equ:bsigsDT} are determined to be $(14.90\pm 0.11)\%$, $(13.04\pm 0.15)\%$ and $(11.76\pm 0.13)\%$ for the three tag modes of $K^{+} \pi^{-}$, $K^{+} \pi^{-} \pi^{0}$, and $K^{+} \pi^{-} \pi^{+} \pi^{-}$, respectively.  In the subsequent analysis the summed energy $E_{\rm{EMC}}$ of all the showers deposited in the electromagnetic calorimeter, excluding those used to reconstruct the $\pi^0$'s from the singly-tagged and signal $D$ decays  is fitted as a discriminating variable. Signals are expected to peak close to zero.

The residual background surviving the selection  is found to be dominated by $D^0\to \pi^0 K_{L}^{0}X$ decays, where $X$ denotes multiple neutral and charged soft pions. The energy deposit of these events is not well described in the generic MC sample as is evident from the discrepancy between data and MC  in Fig.~\ref{fig:mm2}.  This discrepancy is attributed both to the poorly known branching fractions of these decays, and  the imperfect simulation of the interaction between $K_{L}^{0}$ mesons and the  detector material by Geant4~\cite{Achasov:2015xea}. Therefore, the energy deposits of $K_{L}^{0}$ and $X$ are modeled in a data-driven method with control samples as discussed below. 

The energy deposit of  $K_{L}^{0}$  mesons is studied with a control sample selected from a data set of 1,087$\times 10^{6}$ $J/\psi$ decays~\cite{jpsisample}. Two control samples of  $J/\psi \to \phi K^{\pm} \pi^{\mp} K_L^0$ and $J/\psi \to K^{\pm} \pi^{\mp} K_L^0$ are selected with a signal purity of about 99\%. The $K_{L}^{0}$-induced showers within $\degree{10}$ with respect to the $K_{L}^0$ flight direction are counted. The $\degree{10}$ cone requirement helps to suppress the contamination from beam background and bremsstrahlung radiation from charged tracks. According to MC simulations, the unaccounted showers outside the $\degree{10}$ cone are generally less than 50 MeV and on average $75\%$ of the energy deposited by the $K_{L}^0$ can be reconstructed within the cone. 
In order to recover the showers that are unaccounted for, we correct the obtained energy deposit in the control sample with a multi-dimensional reweighting method as described in Ref.~\cite{Rogozhnikov:2016bdp}, according to a correction factor predicted from single-track $K_L^0$ simulations as a function of the $K_L^0$ momentum $P_{K_{L}^0}$, the cosine of the polar angle $\cos\theta_{K_{L}^0}$ and the energy deposit $E_{\rm{EMC}}^{K_{L}^{0}}$.

The energy deposit arising from $X$ in the $D^0\to \pi^0 K_{L}^0 X$ decays is studied in a control sample of $D^0 \to \pi^0 K_{S}^0X,~K_{S}^0\to\pi^+\pi^-$ events. In the $\bar{D}^0$ single-tagged events, we require additional $\pi^0$ and $K_{S}^0$ candidates following the selection criteria in Ref.~\cite{Ablikim:2018gro} to select the control sample. For the sample size under consideration, the $X$ components are expected to be same in $D^0$ decays to $\pi^0 K_L^0 X$ and $\pi^0 K_S^0 X$. The purity of the control sample is improved after subtracting non-$D^0\bar{D}^0$ decay contributions according to the $M_{\rm{BC}}$ sideband events in $\bar{D}^0$ single-tag final states. Those showers that are not associated with  $\bar{D}^0$ single-tag tracks, $\pi^0$ or $K_{S}^0$ decays are summed up to represent the $X$ energy deposit denoted as $E_{\rm{EMC}}^{X}$. 

To obtain a $E_{\rm{EMC}}$ distribution that is representative of the $D^0\to \pi^0 K_{L}^{0}X$ background, we obtain a value for both $E_{\rm{EMC}}^{K_{L}^{0}}$ and $E_{\rm{EMC}}^{X}$ by randomly sampling from the two separate control samples and sum them up. In order to account for differences in the momentum distribution of the $K^0_L$ between the control samples and the decay $D^0\to \pi^0 K_{L}^{0}X$, the control samples are re-weighted to match the $K_{L}^{0}$ momentum distribution in the signal simulation. The signal simulation itself is in turn corrected to account for differences between data and simulation observed in the $K_S^0$ momentum distribution in $D^0\to \pi^0 K_{S}^{0}X$.

The sampling variable is dependent on the kinematic variables of the $K^0_L$ momenta based on MC simulation samples, which are corrected according to the $K_S^0$ momentum difference  observed in the $\pi^0 K_S^0 X$ control sample and simulations.
The resulting $E_{\rm{EMC}}$ distribution for $D^0\to \pi^0 K_{L}^{0}X$ is shown in Fig.~\ref{fig:klpi0xshape}.   

There are two remaining sources of background. One arises from wrongly tagged events from non-$D^0\bar{D}^0$ processes. Their contributions can be estimated with the events in the singly-tagged $M_{\rm{BC}}$ sideband region, which is defined as (1.830, 1.855)$\,\mathrm{GeV}/c^2$. The number of the events in the sideband region is $3134$ and the corresponding scaling factor of the background numbers in the signal region, which is defined as (1.858, 1.874)$\,\mathrm{GeV}/c^2$, and sideband region is found to be $0.611\pm0.001$~\cite{Ablikim:2018gro}, which results in an estimate of $1919\pm34$ background events in data. The other source of contamination is from events with a correct $\bar{D}^0$-tag candidate but without a $K_{L}^{0}$ meson in the $D^0$ decay final states. These $D^0$ decay backgrounds involve decay products with soft kaons and pions, which fail to be reconstructed in the detector. This contribution is modeled with a simulated sample of  $\psi(3770)\to D^{0}\bar{D}{}^{0}$ decays.

\begin{figure}[h!]
\centering
\includegraphics[width=\linewidth]{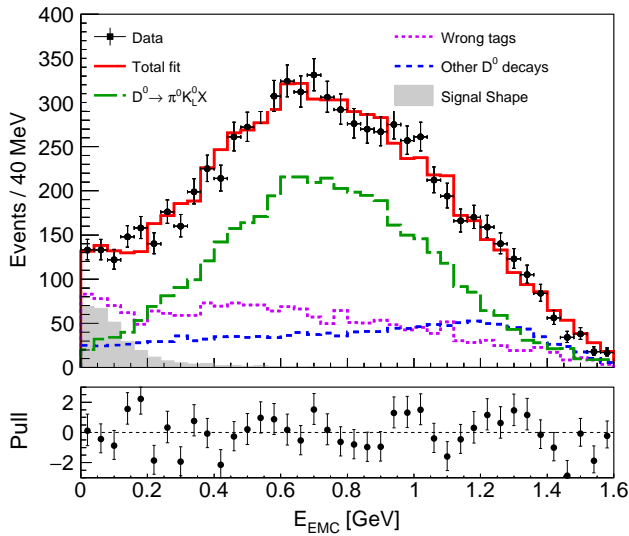}
\caption{Fit to the $E_{\rm{EMC}}$ distribution in data. The black dots with uncertainties represent data and the red solid line shows the total fit. The background components of $D^0 \to K_{L}^{0}\pi^0 X$ decays, wrong $\bar{D}^0$ tags and other $D^0$ decays are marked as green long-dashed line, purple dotted line and blue dashed line, respectively. The gray filled area shows the signal shape, normalized to 20 times the central value of the fit result for visibility.  The bottom panel shows the fit residuals.}
\label{fig:fit_ST}
\end{figure}

Figure~\ref{fig:fit_ST} presents the $E_{\rm{EMC}}$ distribution in data, on which the result of an extended maximum likelihood fit is overlaid. In the fit, the shape of the signal process is modeled according to MC simulations. The yields of all components are free parameters, except for the number of wrongly-tagged $\bar{D}^0$ events, which is fixed to 1919. The fit determines the signal yield $N_{\rm sig}$ to be $14\pm 30$, which is consistent with zero. The reduced $\chi^2$ of the goodness-of-fit test is 1.4. As no signal is observed, an upper limit on the branching fraction of $D^{0} \to \pi^{0} \nu \bar{\nu}$ is estimated after taking into account the systematic uncertainties.

The systematic uncertainties associated with the $\pi^{0}$ reconstruction, the requirement on the number of charged tracks and $\pi^0$ candidates, and the requirement on $M_{\rm{miss}}^2$ are studied with a control sample of $D^0 \to K^{-}\pi^{+}\pi^{0}$ events tagged by the same three $\bar{D}^0$ modes used in this analysis. When imposing our selection requirements, the efficiency differences between data and MC simulations are found to be 2.0\%, 4.0\%, 1.6\% and 0.7\% for the four criteria, respectively. We test the signal model by replacing the $D \to \pi$ form factor with a phase-space model. The resulting efficiency change is found to be 0.5\%, which is assigned as the corresponding systematic uncertainty. To investigate possible bias in the estimate of the signal yield from  the $D^0\to \pi^0 K_{L}^0 X$ modeling, we vary the cone angle to $\degree{15}$ to obtain the $K_{L}^0$ energy deposit and repeat the whole analysis procedure. The change in the signal yield is found to be negligible. 
To determine the size of the uncertainty coming from a wrong estimate of the number of $\bar{D}^0$ wrong-tag events we free this contribution in the fit, with a Gaussian constraint on its uncertainty and find a change on the signal yield of 1.7 compared to the nominal value.  The uncertainty of the branching fraction of $\pi^0\to \gamma\gamma$ is small enough to be neglected~\cite{pdg}. All the systematic uncertainties are summarized in Table~\ref{tab:errDT}. 

\begin{table}[!htb]
\begin{center}
\caption{Summary of systematic uncertainties on the signal yield and detection efficiencies.}
\label{tab:errDT}
\begin{tabular}{lccc} \toprule
Source & Size\\ \midrule
Number of $\pi^0$  &4.0$\%$ \\
$\pi^0$ reconstruction  & 2.0$\%$  \\
Number of charged tracks  &1.6$\%$ \\
$M_{\rm{miss}}^2$ requirement &0.7$\%$ \\
Signal model &0.5$\%$ \\
Wrong-tag background  &1.7\\  
$\pi^0 K_L^{0} X$ background shape & Negligible\\
Branching fraction of $\pi^0\to \gamma\gamma$ &Negligible\\
\bottomrule
\end{tabular}
\end{center}
\end{table}

\begin{figure}[h!]
\centering
\includegraphics[width=1.\linewidth]{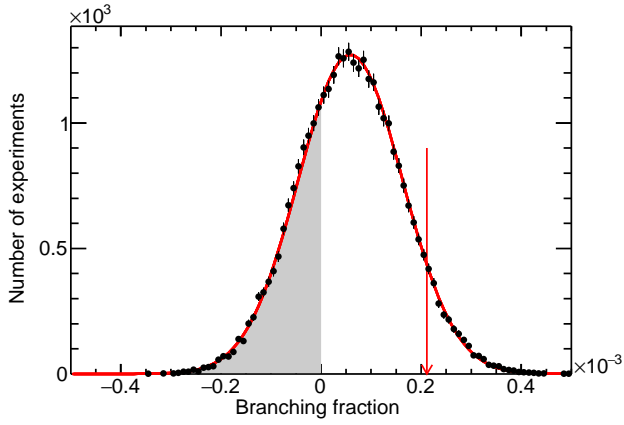}
\caption{Distribution of branching fractions determined from toy samples and a Gaussian fit, which are marked as dots with uncertainties and solid line, respectively. The shaded area represents the non-physical region and the red arrow points to the position of upper limit at the $90\%$ confidence level.}
\label{fig:scan}    
\end{figure}

To set the upper limit of the branching fraction in Eq.~\eqref{equ:bsigsDT} incorporating the systematic uncertainties, we follow the method described in Refs.~\cite{Lees:2011qz,BESIII:2015kvk}. 
An ensemble of 100,000 toy samples is generated as a function of $E_{\rm{EMC}}$ according to the observed data distribution as shown in Fig.~\ref{fig:fit_ST}. In each toy sample, the sample size is randomly sampled from a Poisson distribution with the number of observed events of 7652 in data. A similar fit  to the toy sample is carried out as in data, where different systematic uncertainties are randomly varied.
The contribution of wrong-$\bar{D}^0$-tag backgrounds is fixed to a value constrained by a Gaussian distribution with the central value of 1919 and uncertainty of 34; the involved values in the denominator in Eq.~\eqref{equ:bsigsDT} are randomly assigned according to Gaussian distributions with the nominal values as a mean and their uncertainties corresponding to the standard deviation. The uncertainties associated with the signal-detection efficiencies of the different sources given in Table~\ref{tab:errDT}, i.e. number of $\pi^0$, $\pi^0$ reconstruction, number of charged tracks, $M_{\rm{miss}}^2$ requirement and signal model, are summed up in quadrature. The distribution of the obtained branching fractions in these toy samples is shown in Fig.~\ref{fig:scan}, which follows a Gaussian function as expected. 
Integrating the Gaussian function in the physical region greater than zero, the upper limit of $\mathcal{B}_{\rm sig}$ is determined to be $2.1\times 10^{-4}$ at the $90\%$ confidence level.

In conclusion, we present the first experimental search for the FCNC process $c\to u\nu\bar{\nu}$ through $D^{0} \to \pi^{0} \nu \bar{\nu}$ with $10.6\times10^{6}$ pairs of $D^0\bar{D}^0$ mesons produced by $e^+e^-$ collisions near threshold in the BESIII experiment. Adopting the double tag method and data-driven background modeling, no obvious signal is observed and an upper limit on its decay branching fraction is set to be $2.1\times 10^{-4}$ at 90$\%$ confidence level. This is the world-first experimental constraint in charmed hadron decays to a dineutrino final state. In the absence of sizable long-distance QCD contributions in this process, the result serves as a clean test for $C\!P$ violation in the charm sector, new physics models beyond SM in FCNC decays, as well as for lepton (flavour) universality. 
Our result is lower than the upper limit predicted in Ref.~\cite{Faisel:2020php}, hence, providing constrains on the fermionic coupling strength of leptoquarks to the sterile neutrinos.
In the future, more stringent results will be available based on an anticipated larger $\psi(3770)$ data set of about 20~fb$^{-1}$ at BESIII~\cite{Li:2021iwf,Ablikim:2019hff}.

\acknowledgments
The BESIII Collaboration thanks the staff of BEPCII and the IHEP computing center for their strong support. This work is supported in part by National Key Research and Development Program of China under Nos. 2020YFA0406400, 2020YFA0406300; National Natural Science Foundation of China (NSFC) under Contracts Nos. 11625523, 11635010, 11735014, 11805086, 11822506, 11835012, 11935015, 11935016, 11935018, 11961141012; the Chinese Academy of Sciences (CAS) Large-Scale Scientific Facility Program; the CAS PIFI program; Joint Large-Scale Scientific Facility Funds of the NSFC and CAS under Contracts Nos. U1732263, U1832207; CAS Key Research Program of Frontier Sciences under Contracts Nos. QYZDJ-SSW-SLH003, QYZDJ-SSW-SLH040; 100 Talents Program of CAS; the Fundamental Research Funds for the Central Universities; INPAC and Shanghai Key Laboratory for Particle Physics and Cosmology; ERC under Contract No. 758462; European Union Horizon 2020 research and innovation programme under Contract No. Marie Sklodowska-Curie grant agreement No 894790; German Research Foundation DFG under Contracts Nos. 443159800, Collaborative Research Center CRC 1044, FOR 2359, GRK 2149; Istituto Nazionale di Fisica Nucleare, Italy; Ministry of Development of Turkey under Contract No. DPT2006K-120470; National Science and Technology fund; Olle Engkvist Foundation under Contract No. 200-0605; STFC (United Kingdom); The Knut and Alice Wallenberg Foundation (Sweden) under Contract No. 2016.0157; The Royal Society, UK under Contracts Nos. DH140054, DH160214; The Swedish Research Council; U. S. Department of Energy under Contracts Nos. DE-FG02-05ER41374, DE-SC-0012069.


\end{document}